\documentclass[12pt]{article} 
\usepackage{epsfig,amsmath,graphicx}
\usepackage{color}
\usepackage{amssymb}
\def\tracemumu{T^\mu_{~\mu}}
\begin{document}
\titlepage

\begin{flushright}
OCHA-PP-320\\    April 2014\\
\end{flushright}

\vspace*{1.0cm}

\begin{center}
{\Large \bf 
Production and decay of radion in Randall-Sundrum model at a photon
 collider
} 

\vspace*{1.0cm} 

Gi-Chol Cho$^a$ and Yoshiko Ohno$^b$

\vspace*{0.5cm}
$^a${\em Department of Physics, Ochanomizu University, Tokyo 112-8610, Japan}\\
$^b${\em Graduate School of Humanities and Sciences, 
         Ochanomizu University, Tokyo 112-8610, Japan}\\
\end{center}

\vspace*{1cm}

\baselineskip 18pt
\begin{abstract}
\noindent
A warped extra dimension model predicts an extra scalar particle beyond
 the Standard Model which is called a radion. 
Although interactions of the radion are similar to those of the Higgs boson 
 in the Standard Model, a relatively light radion ($\lesssim 100~{\rm
 GeV}$) is not severely constrained from the Higgs search experiments at
 the LHC.  
In this paper we study discovery potential of the radion at a photon
 collider as an option of ILC. 
Owing to the trace anomaly of the energy-momentum tensor, both a
 production of radion in $\gamma  \gamma$ collision and its decay to
 gluon pair are enhanced sizably.  
We find that the photon collider has a sensitivity for discovering the 
 radion in low-mass region up to $\Lambda_\phi\sim 3~{\rm TeV}$, where
 $\Lambda_\phi$ is a scale parameter which suppresses the interactions
 of radion to the Standard Model particles. 
\end{abstract}

\newpage
\section{Introduction}
A warped extra dimension model proposed by Randall and Sundrum (RS) is 
one of the attractive candidates to solve the gauge hierarchy problem in
the Standard Model (SM) naturally~\cite{Randall:1999ee}. 
The model is given in the five-dimensional space-time where one warped
extra dimension is compactified on the orbifold $S^1/Z_2$.  
The space-time metric is given by 
\begin{align}
 ds^2 = e^{- 2 k y} \eta_{\mu\nu} dx^\mu dx^\nu - dy^2~, 
\end{align}
where $x^\mu (\mu = 0, 1,2, 3$), $y$ and $k$ denote the coordinate of  
four-dimensional space-time, that of a fifth dimension, and the ${\rm
AdS_5}$ curvature, respectively.  
The Minkowski metric is $\eta_{\mu\nu}={\rm diag}(+1,-1,-1,-1)$ and 
$e^{-2ky}$ is called a warp factor. 
Two 3-branes are located at $y=0$ and $\pi r_c$. 
They are called the UV brane ($y=0$) and the IR brane ($y=\pi r_c$). 
In the original RS model all the SM fields are confined on the IR brane
and only graviton is allowed to propagate into the bulk. 
There are some variants of the RS model where some of SM fields
propagate in the extra dimension~\cite{Davoudiasl:1999tf,Davoudiasl:2000wi}. 
%
%%%-----------

%%%-----------
With this setup, 
the four dimensional Planck mass $M_{\rm pl}$ is expressed by 
a fundamental parameter $M$ in the five dimensional Einstein-Hilbert
action as 
\begin{align}
M_{\rm pl}^2=\frac{M^3}{k}(1-e^{-2ky})\label{eq:hr},
\end{align}
As a result, an effective mass parameter in the IR brane is given as 
$M_{\rm pl} e^{-k \pi r_c}$, and the gauge hierarchy problem could be
solved naturally when the distance between UV and IR branes are
stabilized by $kr_c \sim 12$.  
Goldberger and Wise have proposed an attractive mechanism to stabilize
the distance between two branes introducing a bulk scalar field which
has scalar potentials on both branes~\cite{Goldberger:1999uk}.    
Minimizing the scalar potentials on the branes, the distance between two
branes takes appropriate value ($kr_c\sim 12$) without fine-tuning of
the parameters in the scalar potentials. 
In four dimensional effective theory of the original RS model, there are
two new particles beyond the Standard Model. 
One is a spin-2 graviton (and its Kaluza-Klein excitations) and the
other is a scalar-field radion $\phi$ which is a metric fluctuation
along the extra dimension. 
The radion acquires the mass of the order of the electroweak scale due
to the Goldberger-Wise mechanism and it could be a lightest extra particle
in the RS model~\cite{Goldberger:1999uk,Kribs:2006mq}. 
The radion, therefore, is expected to be the first signature of warped
extra dimension models in direct search experiments such as the LHC. 
Phenomenology of the radion can be characterized by two parameters, a 
radion mass $m_\phi$ and a scale parameter $\Lambda_\phi$. 
%%%-----------

%%%-----------
The search experiments of the Higgs boson at the LHC give stringent
constraints on the parameters of the radion. 
As will be shown later, the radion couples to the trace part of the
energy-momentum tensor of the SM.  
Thus it is known that the interactions of the radion to the SM 
particles such as electroweak gauge bosons ($W^\pm,Z$) and fermions are
similar to those of the Higgs boson, except for the scale
parameters in the couplings.  
On the other hand, the interactions of radion to photons and gluons have
additional source from the trace anomaly of the energy-momentum tensor
in addition to the 1-loop contributions from $W$ boson and/or fermions
as in the SM. 
Furthermore, there could be a mixing between the radion and the SM Higgs 
boson through the scalar-curvature mixing term in the 
four-dimensional effective action~\cite{Giudice:2000av, Csaki:2000zn}. 
According to these characteristic features of the radion interactions, 
there have been studied the phenomenological aspects of the radion 
in various colliders~\cite{Dominici:2002jv, Desai:2013pga, Battaglia:2003gb,
Cheung:2003ze,Abbiendi:2004vx, Bae:2001id, Mahanta:2000ci}.  
%%%-----------

%%%-----------
It has been reported that 
the Higgs boson mass $m_h$ is $126.0\pm 0.4{\rm (stat.)}\pm 0.4{\rm
(syst.)}~{\rm GeV}$
at ATLAS~\cite{Aad:2012tfa} and $125.3\pm 0.4{\rm (stat.)} \pm 0.5{\rm
(syst.)}~{\rm GeV}$ at
CMS~\cite{Chatrchyan:2012ufa}, respectively, and there is no signature
of any other scalar particles up to $600~{\rm GeV}$ from the $ZZ$ mode.
Thus one can apply the results of the Higgs search experiments at the 
LHC to constrain the mass and couplings of the radion.   
Recently, 
the bounds on the parameters $m_\phi$ and $\Lambda_\phi$ 
were studied in light of the Higgs boson discovery at the 
LHC using $pp\to h\to \gamma \gamma, ZZ, W^+W^-$~\cite{
Desai:2013pga, Cho:2013mva}. 
It was found that constraint on the parameter $\Lambda_\phi$ for the 
high-mass region of the radion ($m_\phi \lesssim 1~{\rm TeV}$) 
from the $ZZ$ mode is much severe than results in the 
previous studies~\cite{Gunion:2003px, Barger:2011qn}. 
It is, however, pointed out that the radion in low-mass region 
($m_\phi \sim 100~{\rm GeV}$) is not constrained at
the LHC, i.e., the Higgs search in the $\gamma \gamma$ channel at the
LHC is less sensitive to a relatively light radion, since the $\phi \to
gg$ mode dominates over the other decay modes in this region which
suppresses the branching ratio of $\phi \to \gamma \gamma$. 
Then it is worth examining possibilities to search for the radion in the
low-mass region in collider experiments. 
%%%-----------

%%%-----------
In this paper, we study production and decay of the radion at a photon
collider which has been proposed as an option of an $e^+ e^-$ linear
collider such as the ILC~\cite{Adolphsen:2013kya}.  
It has been studied that the photon collider has an advantage to
distinguish the radion produced in the $\gamma \gamma$ collision from
the SM Higgs boson production~\cite{Gunion:2004nx,
Chaichian:2001gr, Cheung:2000rw} supposing the SM Higgs boson is
relatively heavy. 
The production of radion in the light-by-light scattering at the LHC has
been discussed in ref.~\cite{Fichet:2013gsa}.
We show that the decay of radion in low-mass region into gluon pair is a
promissing channel for its discovery at photon collider.
%%%-----------

%%%-----------
This paper is organized as follows. 
In Sec.~2, we briefly review the interactions of radion to SM fields with
emphasis on the difference from those of the SM Higgs boson. 
Production and decay of the radion at the photon collider are discussed
in Sec.~3. 
We show our numerical results in Sec.~4 where a discovery 
potential of the radion at the photon collider is discussed
quantitatively.   
Sec.~5 is devoted to summary and discussion. 

\section{Interactions}

The radion field represents a fluctuation of the distance between the UV 
and IR branes.  
Taking account of a fluctuation along the fifth dimension, the 
metric is written as~\cite{Csaki:2000zn}
\begin{eqnarray}
 ds^2 = e^{-2(ky + F(x,y))}\eta_{\mu\nu}dx^\mu dx^\nu -(1+2F(x,y))^2
	dy^2, 
\label{metric}
\end{eqnarray}
where $F(x,y)$ is a scalar perturbation. 
A canonically normalized radion field $\phi$ is given by~\cite{Csaki:2000zn} 
\begin{eqnarray}
 F(x,y)=\frac{\phi}{\Lambda_\phi} e^{2k(y-\pi r_c)}, 
\end{eqnarray}
where the scale parameter $\Lambda_\phi$ is ${\cal O}({\rm
TeV})$~\cite{Csaki:2000zn, Csaki:1999mp}. 
%%%-----------

%%%-----------
The radion couples to the trace part of the energy-momentum tensor of the SM. 
Then, the interaction Lagrangian of the radion is given by 
\begin{align}
 {\cal L}_{\rm int} =
 \frac{\phi}{\Lambda_\phi} \tracemumu, 
\label{eq:interaction_Lagrangian}
\end{align}
where $\tracemumu$ is the trace of energy-momentum tensor of the SM
which is given as
\begin{eqnarray}
\tracemumu &=& -2 m_W^2 W_\mu^+ W^{- \mu} - m_Z^2 Z_\mu Z^\mu
 + \sum_f m_f \bar{f} f 
 + (2m_h^2 h^2 - \partial_\mu h  \partial^\mu h) 
\nonumber \\
 &+& \frac{\beta_{\rm QED}}{2e} F_{\mu\nu}F^{\mu\nu} 
 + \frac{\beta_{\rm QCD}}{2g_s} G^a_{\mu\nu}G^{a\mu\nu} 
+ \cdots, 
\label{eq:trace_SM_EMtensor}
\\
\beta_{\rm QED}&=&
\left(\frac{1}{16\pi}\right)^2 \left(\frac{19}{6}g_2^3
 -\frac{41}{6}g_Y^3  \right), 
\\
\beta_{\rm QCD}&=&
\left(\frac{1}{16\pi}\right)^2\left(11-\frac{2}{3}n_f \right)
g_s^3, 
\end{eqnarray}
where the first line in r.h.s. of (\ref{eq:trace_SM_EMtensor}) 
is obtained from the energy-momentum tensor of the SM. 
Two terms in the second line of (\ref{eq:trace_SM_EMtensor}) come from
the trace anomaly for photons ($F_{\mu\nu}$) and gluons
($G^a_{\mu\nu},a=1,\cdots, 8$), respectively, and the ellipsis denote the
higher-order terms. 
The number of active quark-flavors is denoted by $n_f$\footnote{
We take $n_f=6$ in our analysis.}. 
We can see from (\ref{eq:trace_SM_EMtensor}) that, except for the trace 
anomaly terms, the interactions of the radion to the SM fields are very
similar to those of the SM Higgs boson. 
The interactions of the radion are, however, suppressed by the scale
parameter $\Lambda_\phi \sim O({\rm TeV})$ which corresponds to the
Higgs vacuum expectation value, $v =246 {\rm GeV}$, in the interactions
of the SM Higgs boson to other SM fields. 
%%%-----------

%%%-----------
It is worth mentioning that, in general, the radion and the Higgs boson
can mix after the electroweak symmetry breaking through the 
scalar-curvature term in the four-dimensional effective
action~\cite{Giudice:2000av, Csaki:2000zn}. 
In our following study, however, we do not consider the 
mixing between the radion and the Higgs boson, since the current
experimental results of the Higgs searches at the LHC tell us that the 
measured branching ratios of the Higgs boson are consistent with those
in the SM~\cite{Aad:2012tfa, Chatrchyan:2012ufa}, so that 
we can neglect the radion-Higgs mixing in a good approximation. 

\section{Production and decay of radion at a photon collider}
%----------------
The production cross section of the radion in the $\gamma\gamma$ 
collision, and the branching fractions are obtained from the interaction 
Lagrangian in eq.~(\ref{eq:interaction_Lagrangian}). 
In the photon collider, the high energy photons are obtained from
electron beams through the inverse Compton scattering. 
The convoluted production cross section with energy distribution of
photon beams is given by~\cite{Chaichian:2001gr, Cheung:2000rw, Cheung:1992jn,
Telnov:1998qj}
\begin{eqnarray}
\sigma(s)=\int_{m_\phi/\sqrt{s}}^{x_{max}}dz\Bigg[
2z\int^{x_{max}}_{z^2/x_{max}}\frac{dx}{x}\ f_\gamma(x)
 f_\gamma\Big(\frac{z^2}{x}\Big)\Bigg]
\times \hat{\sigma}_{\gamma\gamma\rightarrow \phi}(\hat{s}), 
\label{sigma}
\end{eqnarray}
where $\sqrt{s}$ and $\sqrt{\hat{s}}$ represent the center-of-mass energy
of electron pair and $\gamma\gamma$ systems, respectively.  
A momentum fraction of a photon against the electron momentum is denoted
by $x$, and $z$ is defined by $z^2=\hat{s}/s$. 
The production cross section of the radion in the $\gamma\gamma$
annihilation $\hat{\sigma}_{\gamma\gamma\rightarrow \phi}(\hat{s})$ can 
be expressed using the decay rate of $\phi \to \gamma\gamma$ as 
\begin{eqnarray}
\hat{\sigma}_{\gamma\gamma\rightarrow \phi}(\hat{s})
&=& 
 \cfrac{4\pi^2}{\hat{s}}\ \delta(\sqrt{\hat{s}}-m_\phi)\ \Gamma(\phi\rightarrow
 \gamma\gamma)\notag \\[5pt]
&=&
 \cfrac{4\pi^2}{zs\sqrt{s}}\ \delta\bigg(z-\frac{m_\phi}{\sqrt{s}}\bigg)\
 \Gamma(\phi\rightarrow \gamma\gamma)\ .
\label{cs_hat}
\end{eqnarray}
A function $f_\gamma(x)$ in (\ref{sigma}) is the 
unpolarized photon flux from the laser back-scattering
\begin{eqnarray}
f_\gamma(x) = \frac{1}{D(\xi)}\bigg(
1-x+\frac{1}{1-x}-\frac{4x}{\xi(1-x)}+\frac{4x^2}{\xi^2(1-x)^2}
\bigg) ,
\end{eqnarray}
where
\begin{eqnarray}
D(\xi)=
\bigg(
1 - \frac{4}{\xi} - \frac{8}{\xi^2}
\bigg)
\ln(1+\xi) + \frac{1}{2} + \frac{8}{\xi} - \cfrac{1}{2(1+\xi)^2}\ . 
\end{eqnarray}
A parameter $\xi$ is chosen to be $\xi=4.8$, 
then $D(\xi)=1.8$ and $x_{max}=0.83$~\cite{Cheung:1992jn}. 
%%%-----------

\begin{figure}[htbp]
 \begin{minipage}{0.5\hsize}
  \begin{center}
   \includegraphics[scale=0.3]{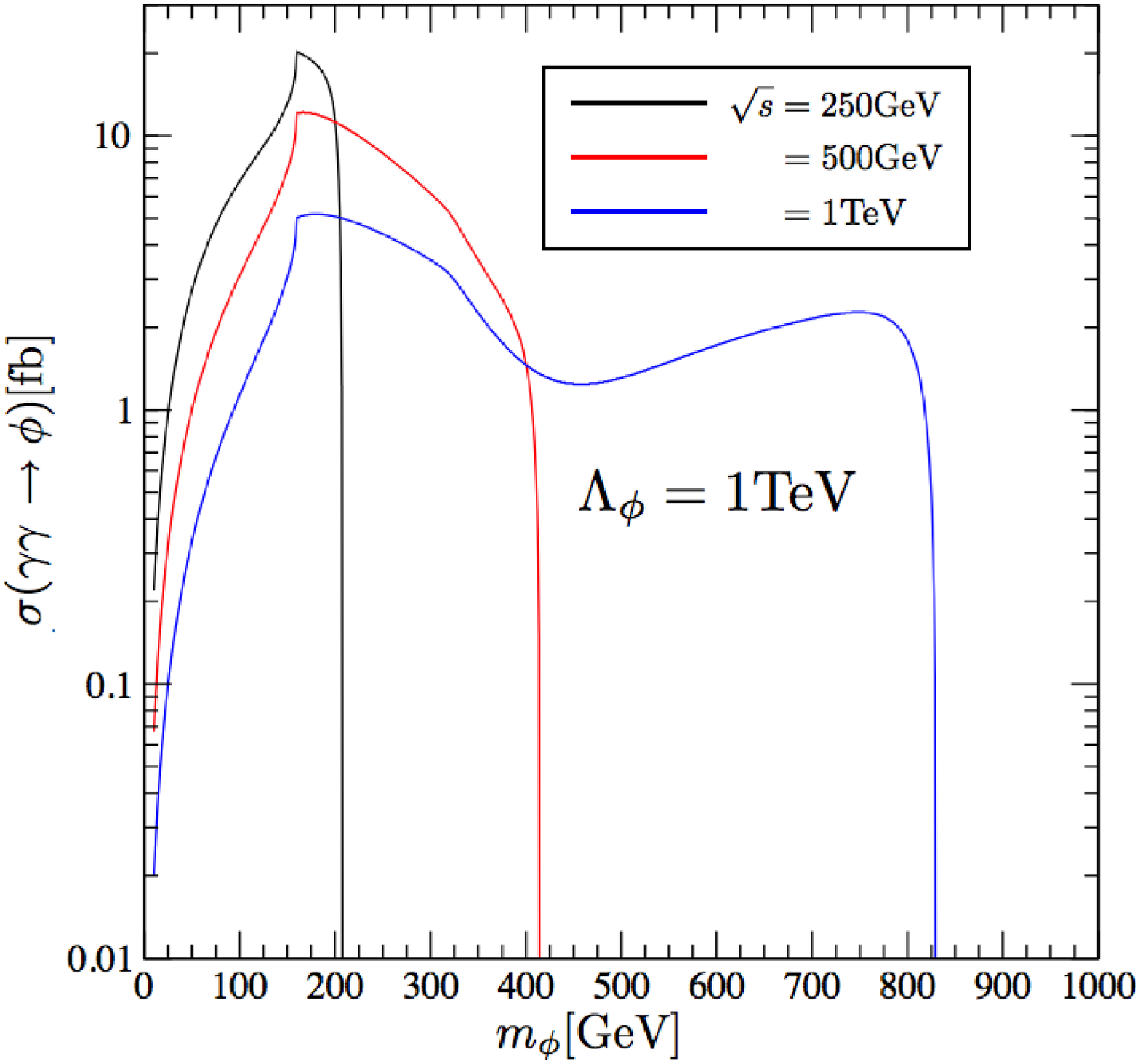}
\quad(a)
  \end{center}
  \label{fig:cs_gam_r_1}
 \end{minipage}
 \begin{minipage}{0.5\hsize}
  \begin{center}
\vspace*{0.1cm}
   \includegraphics[scale=0.3]{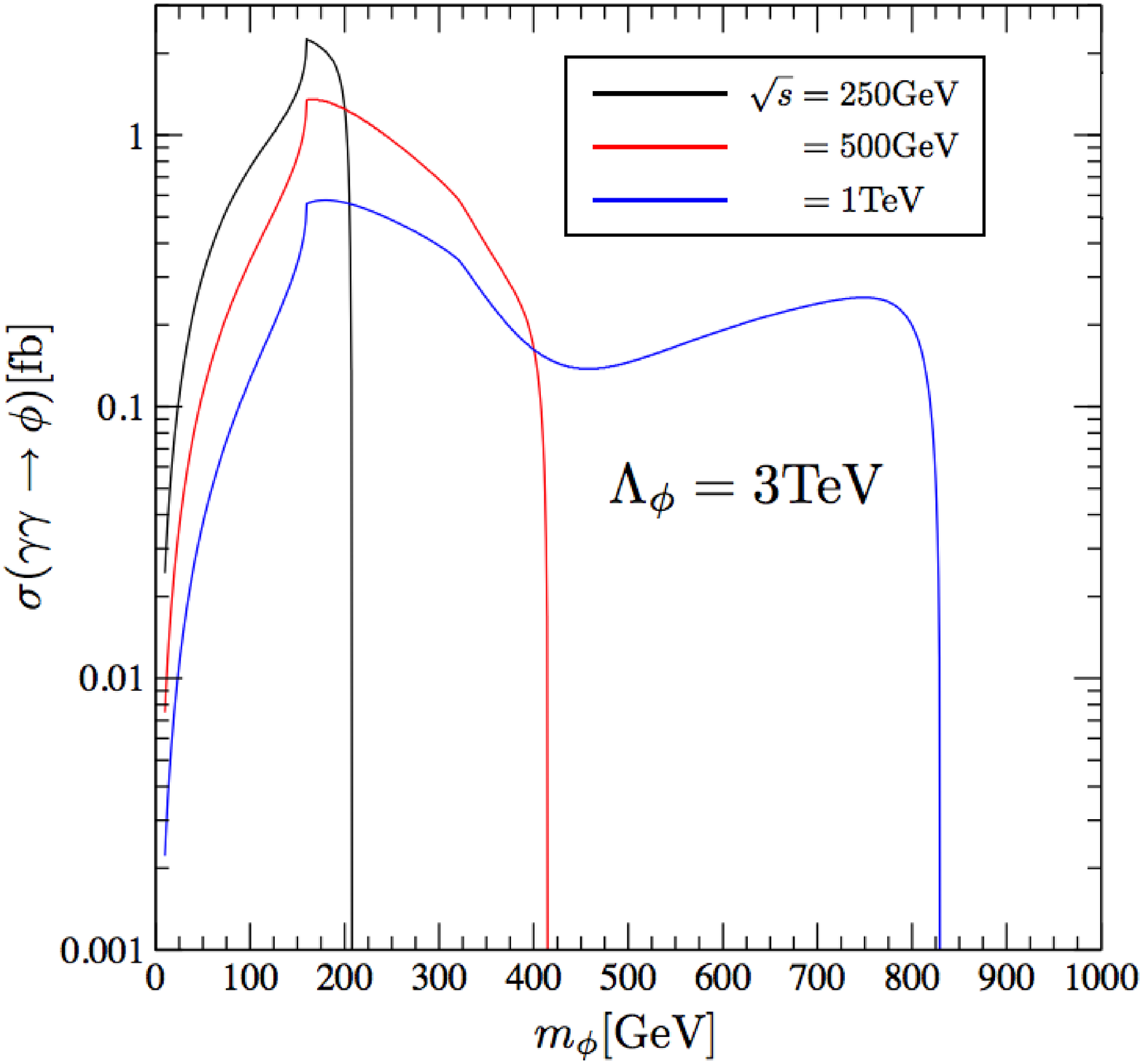}
\quad(b)
  \end{center}
  \label{fig:cs_gam_r_2}
 \end{minipage}
\vspace*{-0.5cm}
  \caption{
The production cross sections of the radion at a photon collider
 given in Eq.~{(\ref{sigma})} as a function of radion mass $m_\phi$ with
 $\Lambda_\phi=1{\rm TeV\ (a)}$ and $\Lambda_\phi=3{\rm TeV\ (b)}$.
 The curves correspond to the electron beam energy 
$\sqrt{s} =
 250{\rm \ GeV}\ ({\rm black}),\ 500{\rm \ GeV}\ ({\rm red})$
 and $1{\rm \ TeV}\ ({\rm blue})$.
 }
  \label{fig:cs}
\end{figure}

%%%-----------
In Fig.~\ref{fig:cs}, the production cross section (\ref{sigma}) 
is given as a function of $m_\phi$ for $\Lambda_\phi=1~{\rm TeV}$ (a) and 
$3~{\rm TeV}$ (b), respectively. 
Three curves in black, red and blue in each figure correspond to
$\sqrt{s}=250~{\rm GeV}$, $500~{\rm GeV}$ and $1~{\rm TeV}$,
respectively.  
For each $\sqrt{s}$, the cross section is enhanced around 
$m_\phi=150-200~{\rm GeV}$.   
Note that the threshold energy of electron collision for the radion
production is somewhat smaller than $m_\phi$ since the radion is produced
via collision of back-scattered photons. 
For example, in Fig.~\ref{fig:cs}~(a), 
the production cross section is larger than $1~{\rm fb}$ 
for $m_\phi \lesssim 400~{\rm GeV}$ ($\sqrt{s}=500~{\rm GeV}$)
and $m_\phi \lesssim 820~{\rm GeV}$ ($\sqrt{s}=1~{\rm TeV}$). 
%%%-----------

\begin{figure}[ht]
\begin{center}
\includegraphics[scale=0.3]{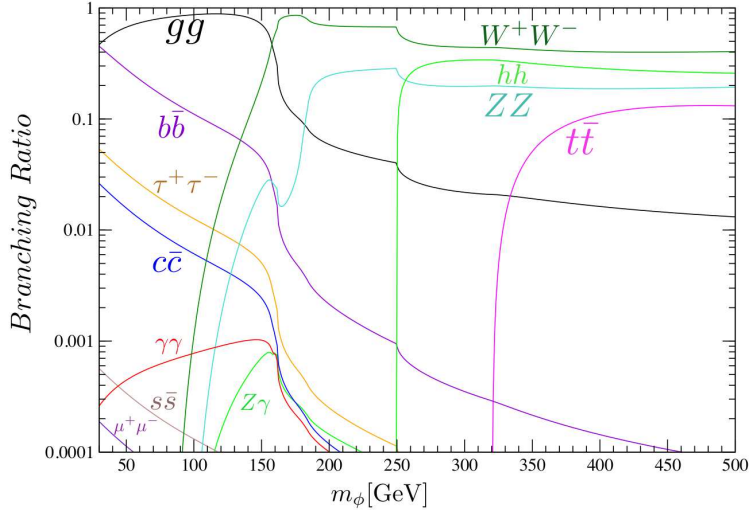}
\end{center}
\vspace*{-0.5cm}
\caption{\label{fig:br}
The branching ratio of radion for each decay modes as a function of
 $m_\phi$. The mass of the SM Higgs boson is fixed at $m_h=125$\ GeV.
}
\label{fig_br}
\end{figure}

%%%-----------
\newpage
The decay widths of the radion to the SM particles are easily 
calculated from eq.~(\ref{eq:interaction_Lagrangian}): 
\begin{eqnarray}
\Gamma(\phi\rightarrow gg) &=&
\frac{\alpha_s^2 m_\phi^3}{32\pi^3\Lambda_\phi^2}
          \left| b_{\rm QCD}+x_t\left\{1+(1-x_t)f(x_t)\right\}\right|^2,
\label{radiongg}
\\
\Gamma(\phi\rightarrow \gamma\gamma) &=& 
\frac{\alpha_{\rm em}^2 m_\phi^3}{256\pi^3\Lambda_\phi^2}
           \Bigg| b_2 + b_Y - \left\{2+3x_W+3x_W(2-x_W)f(x_W)\right\}
						\nonumber \\
           & & \qquad \qquad \qquad \qquad \qquad
 \qquad +\frac{8}{3}x_t\left\{1+(1-x_t)f(x_t)\right\}\Bigg|^2,
\label{radiongamgam}
\\
\Gamma(\phi\rightarrow Z\gamma) &=&
\frac{\alpha_{\rm em}^2 m_\phi^3}{128\pi^3 s_{\rm w}^2
\Lambda_\phi^2}\Bigg(1-\frac{m_Z^2}{m_\phi^2}\Bigg)^3\nonumber\\
& &  \qquad \qquad \qquad 
\times \Bigg| \sum_f N_f \frac{Q_f}{c_{\rm W}} \hat{v}_f \
 A_{1/2}^\phi(x_f,\lambda_f) +A_1^\phi(x_W,\lambda_W)\Bigg|^2,
\label{radionZgamma}
\\
\Gamma(\phi\rightarrow f\bar{f}) &=&
\frac{N_c m_f^2 m_\phi}{8\pi\Lambda_\phi^2}
                             (1-x_f)^{3/2},\\
\Gamma(\phi\rightarrow W^+ W^-) &=&
\frac{m_\phi^3}{16\pi\Lambda_\phi^2}\sqrt{1-x_W}
                            \Big(1-x_W+\frac{3}{4}x_W^2\Big),\\
\Gamma(\phi\rightarrow ZZ) &=&
\frac{m_\phi^3}{32\pi\Lambda_\phi^2}\sqrt{1-x_Z}
                            \Big(1-x_Z+\frac{3}{4}x_Z^2\Big),\\
\Gamma(\phi\rightarrow hh) &=&
\frac{m_\phi^3}{32\pi\Lambda_\phi^2}\sqrt{1-x_h}
                            \Big(1+\frac{1}{2}x_h\Big)^2,
\end{eqnarray}
where $(b_{\rm QCD},b_2,b_Y)=(7, 19/6,-41/6)$. A symbol $f$ denotes all
quarks and leptons. Two variables $x_i$ and
$\lambda_i$ are defined as $x_i = 4m_i^2/m_\phi^2\ (i=t,f,W,Z,h)$ and
$\lambda_i = 4m_i^2/m_Z^2\ (i=f,W)$.
The gauge couplings for QCD and QED are given by $\alpha_s$ and
$\alpha_{\rm em}$, respectively.
The factor $N_f$ is the number of active quark-flavors in the 1-loop
diagrams and
$N_c$ is 3 for quarks and 1 for leptons. 
$Q_f$ and $\hat{v}_f$ denote the electric charge of the
fermion and the reduced vector coupling in the $Zf\bar{f}$ interactions
$\hat{v}_f=2I_f^3-4Q_f s_W^2$, where $I^3_f$ denotes the weak isospin
and $s_W^2\equiv \sin^2{\theta_W},\ c_W^2=1-s_W^2$.
The form factors $A^\phi_{1/2}(x,\lambda)$ and $A^\phi_1
(x,\lambda)$ are given by
\begin{align}
A^\phi_{1/2}(x,\lambda)&=
I_1(x,\lambda)-I_2(x,\lambda)\ ,\\
A^\phi_1(x,\lambda)    &=
c_W\Bigg\{4\Bigg(3-\frac{s_W^2}{c_W^2}\Bigg)I_2(x,\lambda)
+\Bigg[\Bigg(1+\frac{2}{x}\Bigg)\frac{s_W^2}{c_W^2}-\Bigg(5+\frac{2}{x}
\Bigg)\Bigg]I_1(x,\lambda)
\Bigg\}\ \notag.
\end{align}
The functions $I_1(x,\lambda)$ and $I_2(x,\lambda)$ are
\begin{align}
I_1(x,\lambda) &= \frac{x\lambda}{2(x-\lambda)}
    +\frac{x^2\lambda^2}{2(x-\lambda)^2}[f(x^{-1})-f(\lambda^{-1})]
    +\frac{x^2\lambda}{(x-\lambda)^2}[g(x^{-1})-g(\lambda^{-1})]\ ,
\notag\\
I_2(x,\lambda) &=
 -\frac{x\lambda}{2(x-\lambda)}[f(x^{-1})-f(\lambda^{-1})]\ ,
\label{loop}
\end{align}
where the loop functions $f(x)$ and $g(x)$ in (\ref{radiongg}),
(\ref{radiongamgam}) and (\ref{loop})
are given by~\cite{Cheung:2000rw, Djouadi:2005gi}
\begin{align}
 f(x) &=
\begin{cases}
    \left\{\sin^{-1}\Bigg(\cfrac{1}{\sqrt{x}}\Bigg)\right\}^2 & ,\quad x\geq 1 \\[2mm]
    -\cfrac{1}{4}\left(\log\cfrac{1+\sqrt{1-x}}{1-\sqrt{1-x}}-i\pi\right)^2&,
 \quad x < 1 
 \end{cases}
\qquad,\\
\notag\\
 g(x) &=
\begin{cases}
\sqrt{x^{-1}-1}\sin^{-1}\sqrt{x} &, \quad x\leq 1
\\[2mm]
    \cfrac{\sqrt{1-x^{-1}}}{2}\left(\log\cfrac{1+\sqrt{1-x^{-1}}}{1-\sqrt{1-x^{-1}}}-i\pi\right)&,
 \quad x > 1 
 \end{cases}
\qquad.
\end{align} 
%%%------
The branching ratio of the radion for all possible decay modes as a
function of $m_\phi$ is shown in Fig.~\ref{fig_br}. 
The mass of Higgs boson is fixed at $m_h=125~{\rm GeV}$ in the figure. 
We can see from Fig.~\ref{fig_br} that the dominant decay mode is $\phi
\to gg$ for $m_\phi \lesssim 150~{\rm GeV}$
while it is altered by $\phi \to W^+W^-$ for 
$2 m_W \lesssim m_\phi$. 
The decay into $ZZ$ or $hh$ are subdominant for $2m_i \lesssim m_\phi$
($i=Z,h$). 
%%%-----------

%%%-----------
We note here that the dominant decay mode of the radion at low-mass
region is $\phi \to gg$ while that of the SM Higgs boson is $h\to b\bar{b}$. 
This is because that the interaction in the former is enhanced by the
trace anomaly of the energy-momentum tensor as mentioned in a previous
section.  
%-----------------
%-----------------

%%----------------
%%----------------

\section{Numerical analysis}
In this section, we discuss a discovery potential of the radion in
low-mass region ($m_\phi \lesssim 150~{\rm GeV}$) at the photon collider 
focusing on the signal process $\gamma\gamma \to \phi\to gg$, which is
expected to be enhanced by the trace anomaly in both production and
decay processes.  
At the photon collider, the leading background is $\gamma\gamma \to h
\to gg$.
Then we survey the model parameter space $(m_\phi, \Lambda_\phi)$ by
requiring a significance $S/\sqrt{B}>5$ by defining both $S$ and $B$ as 
\begin{eqnarray}
S &=& \int \mathcal{L}_{\it eff}\ dt\ \times \sigma(e e \rightarrow
 \phi)\times {\rm Br}(\phi\rightarrow gg), 
\\
B &=& \int \mathcal{L}_{\it eff}\ dt\ \times \sigma(e e \rightarrow
 h)\times {\rm Br}(h\rightarrow gg). 
\end{eqnarray}
The production cross section $\sigma(e^+e^-\rightarrow \phi)$ is 
identical to that in (\ref{sigma}),  
and $\sigma(e^+e^-\rightarrow h)$ is obtained replacing 
$\hat{\sigma}_{\gamma\gamma\rightarrow \phi}(\hat{s})$ in (\ref{cs_hat}) 
by 
$\hat{\sigma}_{\gamma\gamma\rightarrow h}(\hat{s})$. 
The effective photon luminosity $\int \mathcal{L}_{\it eff}\ dt$ is
assumed to be $1/3$ of the electron luminosity following 
the TESLA technical design report~\cite{Badelek:2001xb}. 
In our numerical analysis, we fix the Higgs boson mass $m_h$ 
at $125~{\rm GeV}$ for simplicity.
%%%-----------

%%%-----------
Although the dominant decay channel is $\phi \to gg$ for $m_\phi
\lesssim 150~{\rm GeV}$, it is altered by $\phi \to VV$ ($V=W^\pm,Z$) 
for $2 m_W \lesssim m_\phi$. 
It is very hard to study the radion production and decay using the $W^+W^-$
channel since $\gamma\gamma\to W^+W^-$ occurs at the tree level in the SM and
it overwhelms the signal process. 
Thus we use the $ZZ$ channel instead of $W^+W^-$, and estimate $S/\sqrt{B}$
assuming that $B$ is dominated by the $h\to ZZ$ channel as a reference
in the high-mass region.   
%%%-----------

\begin{figure}[htbp]
 \begin{minipage}{0.5\hsize}
  \begin{center}
   \includegraphics[scale=0.3]{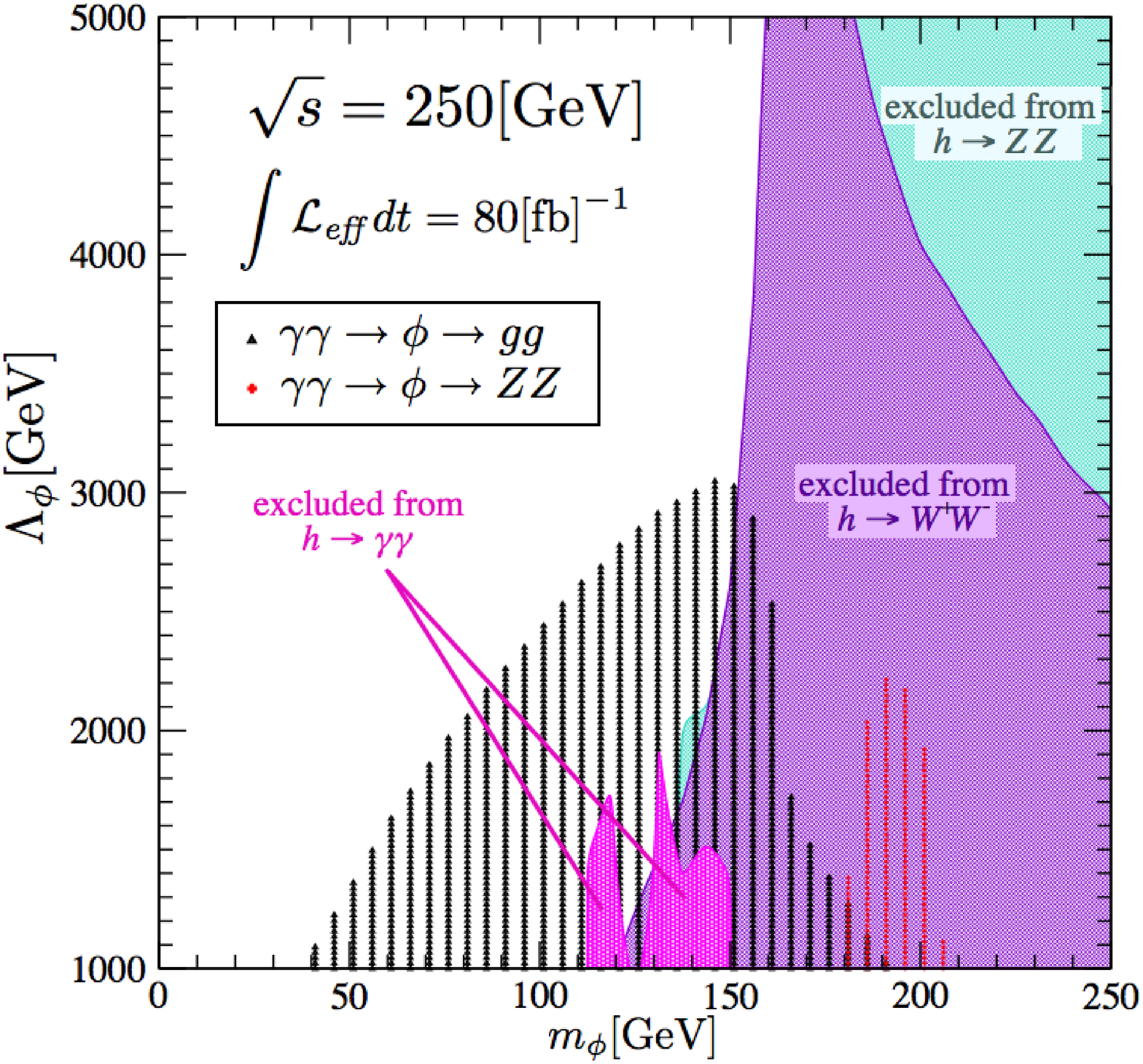}
\quad(a)
  \end{center}
  \label{fig:sig250}
 \end{minipage}
 \begin{minipage}{0.5\hsize}
  \begin{center}
\vspace*{0cm}
   \includegraphics[scale=0.3]{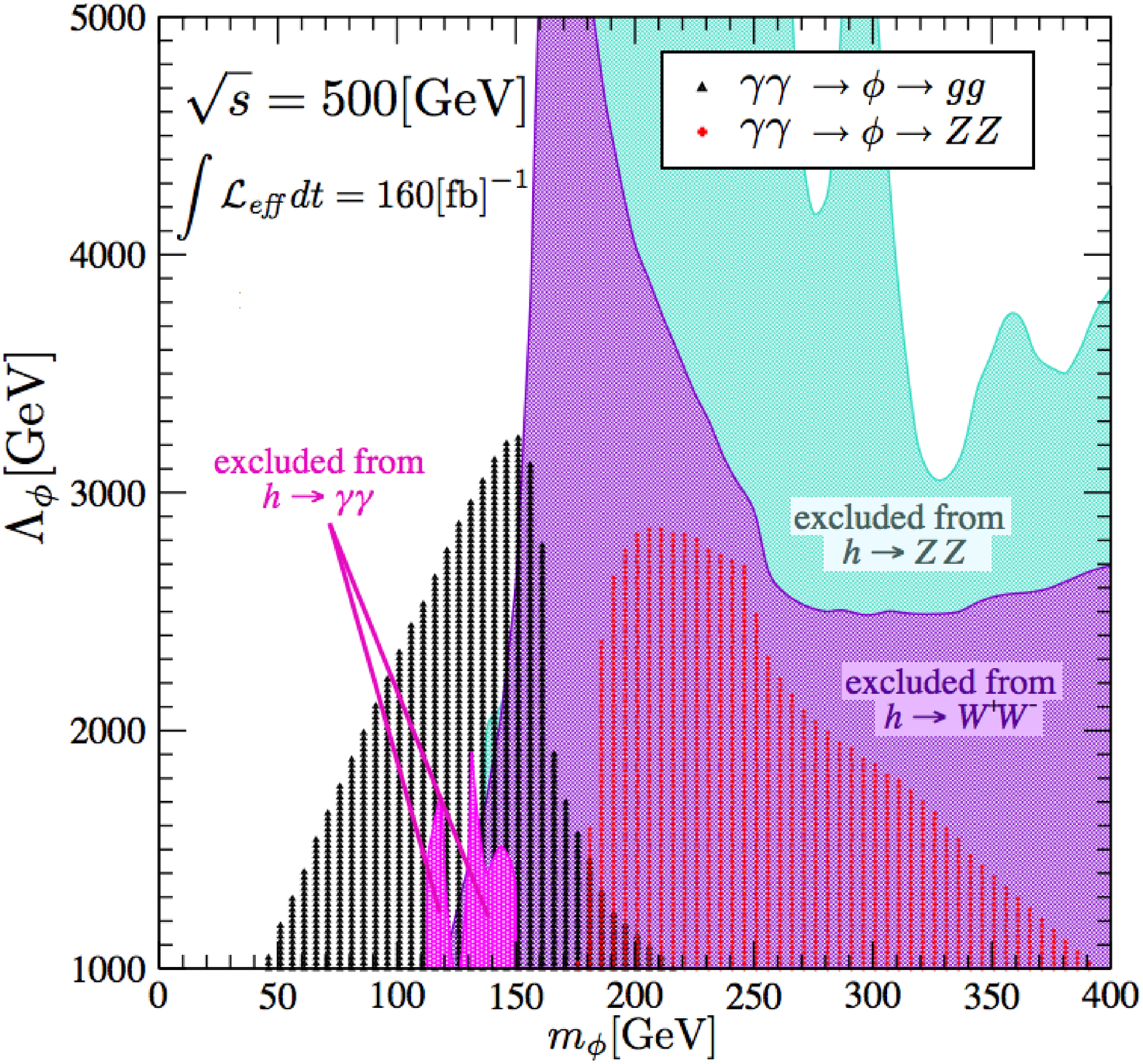}
\quad(b)
  \end{center}
  \label{fig:sig500}
 \end{minipage}

  \begin{center}
\vspace*{0cm}
   \includegraphics[scale=0.3]{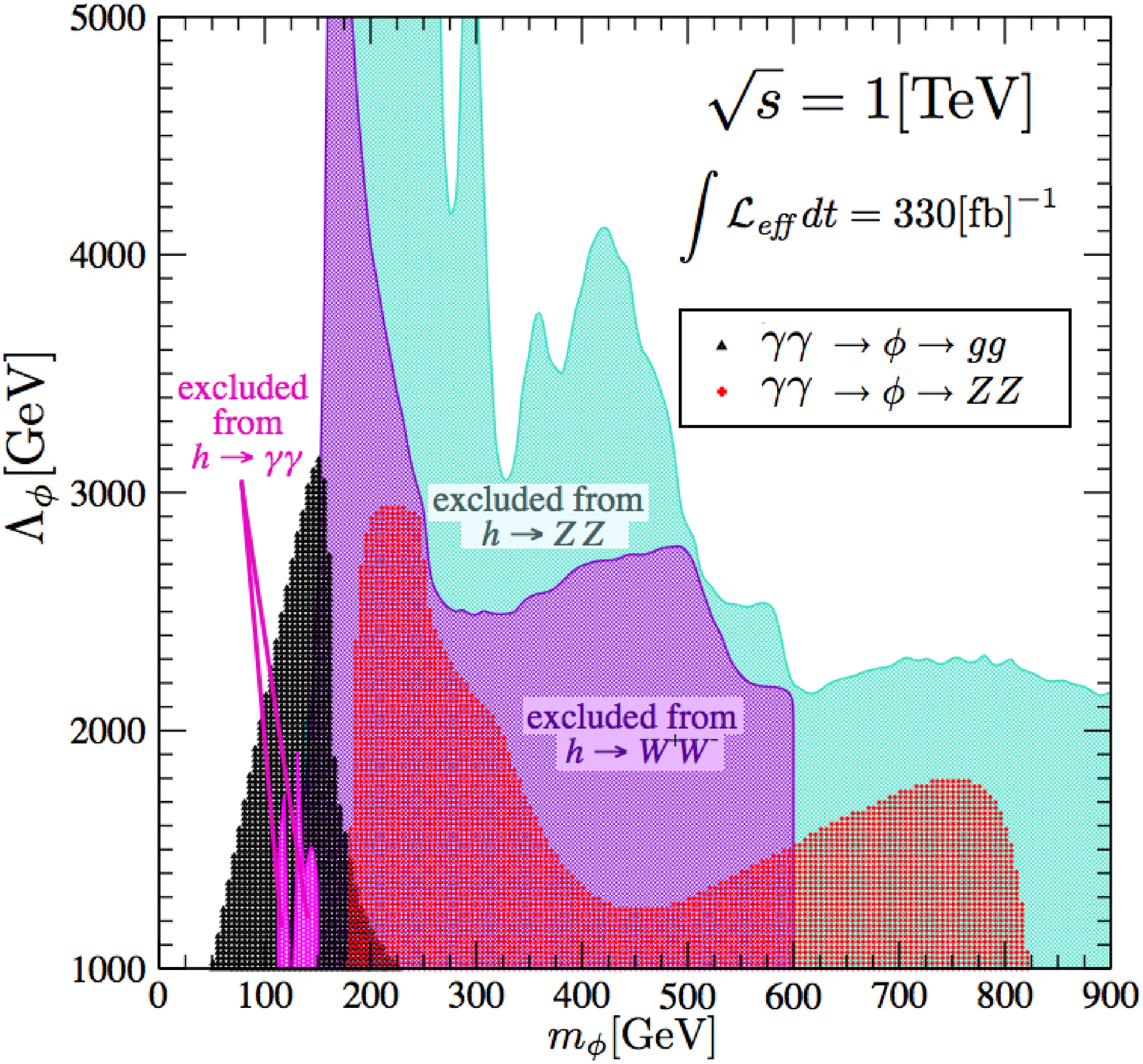}
\\
\quad(c)
  \end{center}
  \label{fig:sig1000}
\vspace*{-0.5cm}
 \caption{
Signal regions on $(m_\phi,\ \Lambda_\phi)$ plane for various
 beam energy of electron
 $\sqrt{s}=250$~GeV~(a), $500$~GeV~(b) and 1~TeV~(c).
 The excluded regions from the recent results from the 
 LHC are also shown~\cite{Cho:2013mva}.
 Black and red regions denote parameter regions where
 signal significance $S/\sqrt{B}>5$ for 
$\gamma \gamma \rightarrow \phi \rightarrow gg$ and
 $\gamma \gamma \rightarrow \phi \rightarrow ZZ$
processes, respectively. Region in turquoise, purple and
 magenta show the $95\%$ CL excluded region from
 $pp\rightarrow h\rightarrow ZZ,\
 pp\rightarrow h\rightarrow W^+W^-$ and $pp\rightarrow h\rightarrow
 \gamma\gamma$ processes at the LHC.}
  \label{fig:sig}
\end{figure}

%%%-----------
In Fig.~\ref{fig:sig}, we show the signal region on
$m_\phi$-$\Lambda_\phi$ plane where $S/\sqrt{B}>5$ is expected for 
various $\sqrt{s}$ and the effective 
integrated luminosity of back-scattered photon 
$\int {\cal L}_{\it eff} dt$.
The expected beam luminosity of ILC is
$(0.75,\, 1.8,\, 3.6) \times 10^{34} {\rm cm^{-2}s^{-1}}$,
for $\sqrt{s}=250~{\rm GeV}$, $500~{\rm GeV}$, $1~{\rm
TeV}$~\cite{Adolphsen:2013kya, Behnke:2013xla}. 
Therefore we used 
$\int {\cal L}_{\it eff} dt=80~{\rm fb}^{-1}$, $160~{\rm fb}^{-1}$ and
$330~{\rm fb}^{-1}$ in Fig.~\ref{fig:sig}~(a), (b) and (c). 
In the figure, black and red regions correspond to  $S/ \sqrt{B} > 5$
 for the $\phi \to gg$ and $\phi \to ZZ$ channels, respectively. 
It is also shown the excluded region on $(m_\phi, \Lambda_\phi)$
plane~\cite{Cho:2013mva} taking account of the Higgs search experiments 
at the LHC~\cite{CMS-PAS-HIG-13-002, CMS-PAS-HIG-13-003,
CMS-PAS-HIG-13-001}.  
Turquoise, purple and magenta regions are excluded from the $pp\to h\to
 ZZ$, $pp\to h\to W^+W^-$ and $pp\to h\to \gamma\gamma$ in the Higgs
 searches at the
LHC. 
%%%-----------

%%%-----------
From the figure, we find that the significance $S/\sqrt{B}$ at least $5$ 
is achieved in both the $gg$ and $ZZ$ channels.  
When $m_\phi \lesssim 150~{\rm GeV}$, $S/\sqrt{B} > 5$ is possible for 
$\Lambda_\phi\lesssim 3~{\rm TeV}$ in (a), (b) and (c). 
On the other hand, the $ZZ$ mode is available only for $180~{\rm
GeV}\lesssim m_\phi$.  
In the $ZZ$ mode, although there are sizable regions with $S/\sqrt{B}>5$
in (b) and (c), these regions are entirely disfavored from the Higgs 
search experiments at the LHC.  
We, therefore, expect that 
the photon collider has a good chance for discovery of the radion with 
$m_\phi \lesssim 150~{\rm GeV}$ but it has no sensitivity to search for 
the radion if the mass is larger than $180~{\rm GeV}$. 
%%%-----------

%%%-----------
We have so far focused on the gluon final state in the radion decay. 
In the experiment, the two gluons in the final state are observed as two
jets which also contain quarks. 
Since $\gamma\gamma\to q\bar{q}$ occurs at the tree level, it is very
crucial to separate the gluon final states from $\gamma\gamma \to jj$. 
The detectors at ILC are aiming to achieve a high efficiency of tagging
the $b$ and $c$ quark flavors~\cite{Behnke:2013xla, Behnke:2013lya}.  
Subtracting $b$ and $c$ jets from the data, and requiring appropriate
kinematical cuts, it is expected to obtain two gluons in dijet data with
a certain efficiency. 
A more quantitative estimation on the background (including 
$\gamma\gamma\to q\bar{q}$) processes is necessary based on the Monte
Carlo 
simulation, which will be done elsewhere~\cite{Ohno:sim}.

\section{Summary}
We have studied production and decay of the radion in the RS 
model at a photon collider as an option of $e^+e^-$ linear collider
(ILC).  
Owing to the trace anomaly of the energy-momentum tensor, the
interactions of the radion to photons and gluons are much enhanced. 
Focusing on the gluon final states in the radion decay, which is a
dominant decay mode in the low-mass region of the radion, 
we investigated the model parameter space where the significance
$S/\sqrt{B}>5$, and found that it could be achieved for $\Lambda_\phi
\lesssim 3~{\rm TeV}$ and $m_\phi \lesssim 150~{\rm GeV}$, without
conflicting the constraints from the LHC experiments.
To be more realistic, it is necessary to estimate both signal and
background (including $\gamma\gamma\to q\bar{q}$) processes using the
Monte Carlo simulation.  
The photon collider could be a good stage to look for the radion in the
low-mass region which LHC experiment does not cover. 

\vspace{1cm}
\newpage
\noindent
{\bf\large Acknowledgements}

\vspace{4mm}
\noindent
The work of G.C.C is supported in part by Grants-in-Aid for Scientific 
Research from the Ministry of
Education, Culture, Sports, Science and Technology (No.24104502) and 
from the Japan Society for the Promotion of Science (No.21244036).  


\begin{thebibliography}{99}
%
%%%% 1 %%%%
\bibitem{Randall:1999ee} 
    L.~Randall and R.~Sundrum,
    ``A Large mass hierarchy from a small extra dimension,''
    Phys.\ Rev.\ Lett.\  {\bf 83}, 3370 (1999)
    [hep-ph/9905221].
    %%CITATION = HEP-PH/9905221;%%

%%%% 2 %%%%
%\cite{Davoudiasl:1999tf}
\bibitem{Davoudiasl:1999tf} 
  H.~Davoudiasl, J.~L.~Hewett and T.~G.~Rizzo,
  ``Bulk gauge fields in the Randall-Sundrum model,''
  Phys.\ Lett.\ B {\bf 473}, 43 (2000)
  [hep-ph/9911262].
  %%CITATION = HEP-PH/9911262;%%
  %404 citations counted in INSPIRE as of 28 Jan 2014

%%%% 3 %%%%
%\cite{Davoudiasl:2000wi}
\bibitem{Davoudiasl:2000wi} 
  H.~Davoudiasl, J.~L.~Hewett and T.~G.~Rizzo,
  ``Experimental probes of localized gravity: On and off the wall,''
  Phys.\ Rev.\ D {\bf 63}, 075004 (2001)
  [hep-ph/0006041].
  %%CITATION = HEP-PH/0006041;%%
  %361 citations counted in INSPIRE as of 28 Jan 2014

%%%% 4 %%%%
%\cite{Goldberger:1999uk}
\bibitem{Goldberger:1999uk} 
  W.~D.~Goldberger and M.~B.~Wise,
  ``Modulus stabilization with bulk fields,''
  Phys.\ Rev.\ Lett.\  {\bf 83}, 4922 (1999)
  [hep-ph/9907447].
  %%CITATION = HEP-PH/9907447;%%
  %937 citations counted in INSPIRE as of 28 Jan 2014

%%%% 5 %%%%
%\cite{Kribs:2006mq}
\bibitem{Kribs:2006mq}
  G.~D.~Kribs,
  ``TASI 2004 lectures on the phenomenology of extra dimensions,''
  hep-ph/0605325.
  %%CITATION = HEP-PH/0605325;%%
  %53 citations counted in INSPIRE as of 29 Jan 2014

%%%% 6 %%%%
\bibitem{Giudice:2000av} 
  G.~F.~Giudice, R.~Rattazzi and J.~D.~Wells,
  ``Graviscalars from higher dimensional metrics and curvature Higgs mixing,''
  Nucl.\ Phys.\ B {\bf 595}, 250 (2001)
  [hep-ph/0002178].
  %%CITATION = HEP-PH/0002178;%%
%
%%%% 7 %%%%
\bibitem{Csaki:2000zn} 
    C.~Csaki, M.~L.~Graesser and G.~D.~Kribs,
    %{\it ``Radion dynamics and electroweak physics,''}
    Phys.\ Rev.\ D {\bf 63}, 065002 (2001)
    [hep-th/0008151].
    %%CITATION = HEP-TH/0008151;%%

%%%% 8 %%%%
%\cite{Dominici:2002jv}
\bibitem{Dominici:2002jv} 
  D.~Dominici, B.~Grzadkowski, J.~F.~Gunion and M.~Toharia,
  ``The Scalar sector of the Randall-Sundrum model,''
  Nucl.\ Phys.\ B {\bf 671}, 243 (2003)
  [hep-ph/0206192].
  %%CITATION = HEP-PH/0206192;%%
  %85 citations counted in INSPIRE as of 31 Jan 2014

%%%% 9 %%%%
%\cite{Desai:2013pga}
\bibitem{Desai:2013pga} 
  N.~Desai, U.~Maitra and B.~Mukhopadhyaya,
  ``An updated analysis of radion-higgs mixing in the light of LHC data,''
  arXiv:1307.3765.
  %%CITATION = ARXIV:1307.3765;%%
  %4 citations counted in INSPIRE as of 12 Mar 2014

%%%% 10 %%%%
%\cite{Battaglia:2003gb}
\bibitem{Battaglia:2003gb} 
  M.~Battaglia, S.~De Curtis, A.~De Roeck, D.~Dominici and J.~F.~Gunion,
  ``On the complementarity of Higgs and radion searches at LHC,''
  Phys.\ Lett.\ B {\bf 568}, 92 (2003)
  [hep-ph/0304245].
  %%CITATION = HEP-PH/0304245;%%
  %23 citations counted in INSPIRE as of 31 Jan 2014

%%%% 11 %%%%
%\cite{Cheung:2003ze}
\bibitem{Cheung:2003ze} 
  K.~Cheung, C.~S.~Kim and J.~-h.~Song,
  ``A Probe of the radion Higgs mixing in the Randall-Sundrum model at e+ e- colliders,''
  Phys.\ Rev.\ D {\bf 67}, 075017 (2003)
  [hep-ph/0301002].
  %%CITATION = HEP-PH/0301002;%%
  %15 citations counted in INSPIRE as of 31 Jan 2014

%%%% 12 %%%%
%\cite{Abbiendi:2004vx}
\bibitem{Abbiendi:2004vx} 
  G.~Abbiendi {\it et al.}  [OPAL Collaboration],
  ``Search for radions at LEP2,''
  Phys.\ Lett.\ B {\bf 609}, 20 (2005)
  [Erratum-ibid.\ B {\bf 637}, 374 (2006)]
  [hep-ex/0410035].
  %%CITATION = HEP-EX/0410035;%%
  %5 citations counted in INSPIRE as of 31 Jan 2014

%%%% 13 %%%%
%\cite{Bae:2001id}
\bibitem{Bae:2001id} 
  S.~Bae, P.~Ko, H.~S.~Lee and J.~Lee,
  ``Radion phenomenology in the Randall-Sundrum scenario,''
  hep-ph/0103187
  %%CITATION = HEP-PH/0103187;%%
  %12 citations counted in INSPIRE as of 31 Jan 2014
%\cite{Bae:2000pk}
%\bibitem{Bae:2000pk} 
 and S.~Bae, P.~Ko, H.~S.~Lee and J.~Lee,
  ``Phenomenology of the radion in Randall-Sundrum scenario at colliders,''
  Phys.\ Lett.\ B {\bf 487}, 299 (2000)
  [hep-ph/0002224].
  %%CITATION = HEP-PH/0002224;%%
  %72 citations counted in INSPIRE as of 31 Jan 2014

%%%% 14 %%%%
%\cite{Mahanta:2000ci}
\bibitem{Mahanta:2000ci} 
  U.~Mahanta and A.~Datta,
  ``Production of light stabilized radion at high-energy hadron collider,''
  Phys.\ Lett.\ B {\bf 483}, 196 (2000)
  [hep-ph/0002183].
  %%CITATION = HEP-PH/0002183;%%
  %56 citations counted in INSPIRE as of 31 Jan 2014

%%%% 15 %%%%
\bibitem{Aad:2012tfa} 
  G.~Aad {\it et al.}  [ATLAS Collaboration],
  ``Observation of a new particle in the search for
   the Standard Model Higgs boson with the ATLAS detector at the LHC,''
  Phys.\ Lett.\ B {\bf 716}, 1 (2012)
  [arXiv:1207.7214 [hep-ex]].
 %%CITATION = ARXIV:1207.7214;%%

%%%% 16 %%%%
%
\bibitem{Chatrchyan:2012ufa} 
  S.~Chatrchyan {\it et al.}  [CMS Collaboration],
 ``Observation of a new boson at a mass of 125 GeV
  with the CMS experiment at the LHC,''
  Phys.\ Lett.\ B {\bf 716}, 30 (2012)
  [arXiv:1207.7235 [hep-ex]].
 %%CITATION = ARXIV:1207.7235;%%

%%%% 17 %%%%
\bibitem{Cho:2013mva}
  G.~-C.~Cho, D.~Nomura and Y.~Ohno,
  ``Constraints on radion in a warped extra dimension model from Higgs
	boson searches at the LHC,''
  Mod.\ Phys.\ Lett.\ A {\bf 28}, 1350148 (2013)
  [arXiv:1305.4431 [hep-ph]].
  %%CITATION = ARXIV:1305.4431;%%
  %2 citations counted in INSPIRE as of 24 Dec 2013
%
%%%% 18 %%%%
%\cite{Gunion:2003px}
\bibitem{Gunion:2003px} 
  J.~F.~Gunion, M.~Toharia and J.~D.~Wells,
  ``Precision electroweak data and the mixed Radion-Higgs sector of warped extra dimensions,''
  Phys.\ Lett.\ B {\bf 585}, 295 (2004)
  [hep-ph/0311219].
  %%CITATION = HEP-PH/0311219;%%
  %43 citations counted in INSPIRE as of 12 Mar 2014

%%%% 19 %%%%
%\cite{Barger:2011qn}
\bibitem{Barger:2011qn} 
  V.~Barger and M.~Ishida,
  ``Randall-Sundrum Reality at the LHC,''
  Phys.\ Lett.\ B {\bf 709}, 185 (2012)
  [arXiv:1110.6452 [hep-ph]].
  %%CITATION = ARXIV:1110.6452;%%
  %19 citations counted in INSPIRE as of 12 Mar 2014

%%%% 20 %%%%
%\cite{Adolphsen:2013kya}
\bibitem{Adolphsen:2013kya} 
  C.~Adolphsen, M.~Barone, B.~Barish, K.~Buesser, P.~Burrows, J.~Carwardine, J.~Clark and Hélèn.~M.~Durand {\it et al.},
  ``The International Linear Collider Technical Design Report - Volume 3.II: Accelerator Baseline Design,''
  arXiv:1306.6328 [physics.acc-ph].
  %%CITATION = ARXIV:1306.6328;%%
  %18 citations counted in INSPIRE as of 12 Mar 2014

%%%% 21 %%%%
%\cite{Gunion:2004nx}
\bibitem{Gunion:2004nx} 
  J.~F.~Gunion,
  ``The Need for a photon-photon collider in addition to LHC \& ILC for
 unraveling the scalar sector of the Randall-Sundrum model,''
  hep-ph/0410379.
  %%CITATION = HEP-PH/0410379;%%
  %5 citations counted in INSPIRE as of 31 Jan 2014

%%%% 22 %%%%
%\cite{Chaichian:2001gr}
\bibitem{Chaichian:2001gr} 
  M.~Chaichian, K.~Huitu, A.~Kobakhidze and Z.~H.~Yu,
  ``Radions in a gamma gamma collider,''
  Phys.\ Lett.\ B {\bf 515}, 65 (2001)
  [hep-ph/0106077].
 %%CITATION = HEP-PH/0106077;%%
  %11 citations counted in INSPIRE as of 31 Jan 2014

%%%% 23 %%%%
%cite{Cheung:2000rw}
\bibitem{Cheung:2000rw} 
  K.~-m.~Cheung,
  ``Phenomenology of radion in Randall-Sundrum scenario,''
  Phys.\ Rev.\ D {\bf 63}, 056007 (2001)
  [hep-ph/0009232].
  %%CITATION = HEP-PH/0009232;%%
  %95 citations counted in INSPIRE as of 09 Jan 2014

%%%%%%%%%%%%%% 23plus %%%%%%%%%%%%%%%%%

%\cite{Fichet:2013gsa}
\bibitem{Fichet:2013gsa} 
  S.~Fichet, G.~von Gersdorff, O.~Kepka, B.~Lenzi, C.~Royon and M.~Saimpert,
  ``Probing new physics in diphoton production with proton tagging at the Large Hadron Collider,''
  arXiv:1312.5153 [hep-ph].
  %%CITATION = ARXIV:1312.5153;%%
  %1 citations counted in INSPIRE as of 08 Apr 2014

%%%% 24 %%%%
%\cite{Csaki:1999mp}
\bibitem{Csaki:1999mp} 
  C.~Csaki, M.~Graesser, L.~Randall and J.~Terning,
  ``Cosmology of brane models with radion stabilization,''
  Phys.\ Rev.\ D {\bf 62}, 045015 (2000)
  [hep-ph/9911406].
  %%CITATION = HEP-PH/9911406;%%
  %475 citations counted in INSPIRE as of 15 Jan 2014

%%%% 25 %%%%
%\cite{Cheung:1992jn}
\bibitem{Cheung:1992jn} 
  K.~-m.~Cheung,
  ``Associated production of intermediate Higgs or $Z$ boson with $t \bar{t}$ pair in $\gamma \gamma$ collisions,''
  Phys.\ Rev.\ D {\bf 47}, 3750 (1993)
  [hep-ph/9211262].
  %%CITATION = HEP-PH/9211262;%%
  %66 citations counted in INSPIRE as of 09 Jan 2014

%%%% 26 %%%%
%\cite{Telnov:1998qj}
\bibitem{Telnov:1998qj} 
  V.~I.~Telnov,
  ``Physics goals and parameters of photon colliders,''
  Int.\ J.\ Mod.\ Phys.\ A {\bf 13}, 2399 (1998)
  [hep-ex/9802003].
  %%CITATION = HEP-EX/9802003;%%
  %35 citations counted in INSPIRE as of 09 Jan 2014

%%%% 28 %%%%

%\cite{Djouadi:2005gi}
\bibitem{Djouadi:2005gi} 
  A.~Djouadi,
  ``The Anatomy of electro-weak symmetry breaking. I: The Higgs boson in the standard model,''
  Phys.\ Rept.\  {\bf 457}, 1 (2008)
  [hep-ph/0503172].
  %%CITATION = HEP-PH/0503172;%%
  %758 citations counted in INSPIRE as of 17 Apr 2014


%%%% 29 %%%%
%\cite{Badelek:2001xb}
\bibitem{Badelek:2001xb} 
  B.~Badelek {\it et al.}  [ECFA/DESY Photon Collider Working Group Collaboration],
  ``TESLA: The Superconducting electron positron linear collider with an integrated X-ray laser laboratory. Technical design report. Part 6. Appendices. Chapter 1. Photon collider at TESLA,''
  Int.\ J.\ Mod.\ Phys.\ A {\bf 19}, 5097 (2004)
  [hep-ex/0108012].
  %%CITATION = HEP-EX/0108012;%%
  %215 citations counted in INSPIRE as of 24 Dec 2013

%%%% 30 %%%%
%\cite{Behnke:2013xla}
\bibitem{Behnke:2013xla} 
  T.~Behnke, J.~E.~Brau, B.~Foster, J.~Fuster, M.~Harrison, J.~M.~Paterson, M.~Peskin and M.~Stanitzki {\it et al.},
  ``The International Linear Collider Technical Design Report - Volume 1: Executive Summary,''
  arXiv:1306.6327 [physics.acc-ph].
  %%CITATION = ARXIV:1306.6327;%%
  %26 citations counted in INSPIRE as of 30 Jan 2014

%%%% 31 %%%%
%\cite{Behnke:2013lya}
\bibitem{Behnke:2013lya} 
  T.~Behnke, J.~E.~Brau, P.~N.~Burrows, J.~Fuster, M.~Peskin, M.~Stanitzki, Y.~Sugimoto and S.~Yamada {\it et al.},
  ``The International Linear Collider Technical Design Report - Volume 4: Detectors,''
  arXiv:1306.6329 [physics.ins-det].
  %%CITATION = ARXIV:1306.6329;%%
  %27 citations counted in INSPIRE as of 12 Mar 2014

%%%% 32 %%%%
%
\bibitem{CMS-PAS-HIG-13-002}
    CMS collaboration, ``Properties of the Higgs-like boson in the
    decay $H \to ZZ \to 4 \ell$ in $pp$ collisions at $\sqrt{s}=7$
    and 8 TeV'', CMS PAS HIG-13-002 (2013).
%
%%%% 33 %%%%
\bibitem{CMS-PAS-HIG-13-003}
    CMS collaboration, ``Update on the search for the standard model
    Higgs boson in $pp$ collisions at the LHC decaying to $W^+W^-$
    in the fully leptonic final state'', CMS PAS HIG-13-003 (2013).
%
%%%% 34 %%%%
\bibitem{CMS-PAS-HIG-13-001}
    CMS collaboration, ``Updated measurements of the Higgs boson 
    at 125 GeV in the two photon decay channel'', CMS PAS HIG-13-001 (2013).
%
%%%% 35 %%%%
\bibitem{Ohno:sim}
    Y.~Ohno, in progress. 

\end{thebibliography}
\end{document}